\documentclass[submission,copyright,creativecommons]{eptcs}

\providecommand{\event}{PxTP 2021} % Name of the event you are submitting to
\usepackage{breakurl}             % Not needed if you use pdflatex only.
\usepackage{underscore}           % Only needed if you use pdflatex.
\usepackage{todonotes}

\usepackage[frozencache=true]{minted}
\usepackage{microtype}
\microtypesetup{final,tracking=true,kerning=true,spacing=true,stretch=10,shrink=10,letterspace=25}
\microtypecontext{spacing=nonfrench}

\def\orcid#1{\smash{\href{http://orcid.org/#1}{\protect\raisebox{-1.25pt}{\protect\includegraphics{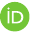}}}}}

\title{Alethe: Towards a Generic SMT Proof Format\\ (extended abstract)}
\author{%
Hans-Jörg Schurr
\orcid{0000-0002-0829-5056}
\institute{University of Lorraine, CNRS, Inria, and LORIA, Nancy, France}
\email{hans-jorg.schurr@inria.fr}
\and
Mathias Fleury
\orcid{0000-0002-1705-3083}
\institute{Johannes Kepler University Linz, Austria}
\email{mathias.fleury@jku.at}
\and
Haniel Barbosa
\orcid{0000-0003-0188-2300}
\institute{Universidade Federal de Minas Gerais, Belo Horizonte, Brazil}
\email{hbarbosa@dcc.ufmg.br}
\and
Pascal Fontaine
\orcid{0000-0003-4700-6031}
\institute{University of Liège, Belgium}
\email{pascal.fontaine@uliege.be}
}

\def\titlerunning{Towards a Generic SMT Proof Format}
\def\authorrunning{Schurr, Fleury, Barbosa, and Fontaine}

\begin{document}
\maketitle

\begin{abstract}
  The first iteration of the proof format used by the SMT solver veriT was
  presented ten years ago at the first PxTP workshop.
  Since then the format has matured. veriT proofs are used
  within multiple applications, and other solvers generate proofs in the
  same format.
  We would now like to gather feedback from the community to guide future
  developments.
  Towards this, we review the history of the format, present
  our pragmatic approach to develop the format, and
  also discuss problems that might arise when other solvers use the format.
\end{abstract}

\noindent
Over the years the production of machine-consumable formal proofs of
unsatisfiability from SMT solvers~\cite{Barrett2018} has attracted significant
attention~\cite{barrett-2015}.
Such proofs enable users to certify unsatisfiability results similarly to how
satisfiable results may be certified via models.
However, a major difficulty that SMT proof formats
must address is the complex and
heterogeneous nature of SMT solvers: a SAT solver drives
multiple, often very
different, theory solvers; instantiation procedures
generate ground instances; and heavily theory-dependent simplification techniques
ensure that the solvers are fast in practice.
Moreover, how each of these components works internally can also differ from
solver to solver.
As a testament to these challenges proof formats for SMT solvers have been
mostly restricted to individual solvers and no standard has emerged.
To be adopted by several solvers, an SMT proof format must be carefully designed
to accommodate needs of specific
solvers. This will require repeated refinement and generalization.

The basis for our efforts in this field is the proof format implemented by the
SMT solver veriT~\cite{verit} that is now mature and used by multiple systems.
To further improve the format, as well as to accommodate not only the reasoning
of the SMT solver veriT but also of other solvers, we are currently extending
the format and developing better tooling, such as an independent proof checker.
To facilitate this effort and overall usage, we are also writing a full
specification.
To emphasize the independence of the format we are baptizing it
\emph{Alethe}.\footnote{Alethe is a genus of small birds and the name resembles \emph{aletheia},
the Greek word for truth.}
We do not presume to propose a standard format a priori. Instead we believe that
Alethe, together with its tooling, can provide a basis for further discussions
on how to achieve a format to be used by multiple solvers.

\section{The State of Alethe}

Alethe combines two major ideas
whose roots reach back ten years to the first PxTP workshop in
2011~\cite{besson-2011,deharbe-2011}.
It was proposed as
an easy-to-produce format with a term language very close to
SMT-LIB~\cite{SMTLIB}, the standard input language of SMT solvers, and rules
with a varying level of granularity, allowing implicit proof steps in the proof
and thus relying on powerful proof checkers capable of filling the gaps.
Since then the format has been refined and extended~\cite{barbosa-2019}. It is now
mature, supports coarse- and fine-grained proof steps capturing
SMT solving for the SMT-LIB logic UFLIRA\footnote{That is the logic for
  problems containing a mix of any of quantifiers,
  uninterpreted
  functions, and linear arithmetic.} and can be reconstructed by the proof
assistants Coq~\cite{Armand2011,SMTCoq} and
Isabelle~\cite{fleury-2019,schurr-2021}.
In particular, the integration with Coq was also used as a bridge for the
reconstruction of proofs from the SMT solver CVC4~\cite{Barrett2011} in
Coq, where its proofs in the LFSC format~\cite{Stump2013} were
first translated into the veriT format before reconstruction.
Finally, the format will also be natively supported in the upcoming cvc5.
solver\footnote{\url{https://cvc4.github.io/2021/04/02/cvc5-announcement.html}}

On the one hand, Alethe uses a
natural-deduction style calculus driven mostly by resolution~\cite{besson-2011}.
To handle
first-order reasoning, dedicated quantifier instantiation rules are used~\cite{deharbe-2011}.
On the other hand, it
implements novel ideas to express reasoning typically used for processing,
such as Skolemization, renaming of variables, and other manipulations
of bound variables~\cite{barbosa-2019}.
While the format was always inspired by the SMT-LIB language,
we recently~\cite{fleury-2019} changed
the syntax of Alethe to closely resemble the command structure
used within SMT-LIB. When possible Alethe uses existing SMT-LIB features,
such as the \texttt{define-fun} command to define constants and the \texttt{:named}
annotation to implement term sharing.

The following proof fragment gives a taste of the format. The
fragment first renames the bound variable in the term $\exists x.\,f(x)$ from
$x$ to $\mathit{vr}$ and then skolemizes the quantifier. A proof is a
list of commands.  The \emph{assume} command introduces an assumption, \emph{anchor}
starts a subproof, and \emph{step} denotes an ordinary proof step. Steps are
annotated with an identifier, a rule, and premises. The SMT-LIB command \emph{define-fun}
defines a function.
The rule \emph{bind} used by step \texttt{t1} performs the renaming
of the bound variable. It uses a subproof (Steps \texttt{t1.t1} and
\texttt{t1.t2}). The subproof uses a context to denote that \texttt{x}
is equal to \texttt{vr} within the subproof. The \texttt{anchor} command
starts the subproof and introduces the context.
The \texttt{bind} rule does not only make it possible to
rename bound variables, but within the subproof it is possible
to simplify the formula as done during preprocessing.
The steps \texttt{t2} and \texttt{t3} use resolution to finish the
renaming.
In step
\texttt{t4} the bound variable is skolemized. Skolemization uses
the choice binder $\epsilon$ and derives $f(\epsilon\mathit{vr}.\, f(\mathit{vr}))$
from $\exists \mathit{vr}.\, f(\mathit{vr})$. To simplify the reconstruction
the choice term is introduced as a defined constant (by \texttt{define-fun}). Finally, resolution
is used again to finish the proof.
\begin{minted}{smtlib2.py -x}
  (assume a0 (exists ((x A)) (f x)))
  (anchor :step t1 :args (:= x vr))
  (step t1.t1 (cl (= x vr)) :rule cong)
  (step t1.t2 (cl (= (f x) (f vr))) :rule cong)
  (step t1 (cl (= (exists ((x A)) (f x))
                  (exists ((vr A)) (f vr)))) :rule bind)
  (step t2 (cl (not (= (exists ((vr A)) (f x))
                       (exists ((vr A)) (f vr))))
               (not (exists ((vr A)) (f x)))
               (exists ((vr A)) (f vr))) :rule equiv_pos1)
  (step t3 (cl (exists ((vr A)) (f vr))) :premises (a0 t1 t2) :rule resolution)
  (define-fun X () A (choice ((vr A)) (f vr)))
  (step t4 (cl (= (exists ((vr A)) (f vr)) (f X))) :rule sko_ex)
  (step t5 (cl (not (= (exists ((vr A)) (f vr)) (f X)))
               (not (exists ((vr A)) (f vr)))
               (f X)) :rule equiv_pos1)
  (step t6 (cl (f X)) :premises (t3 t4 t5) :rule resolution)
\end{minted}

The output of Alethe proofs from veriT  has now reached a certain level of maturity.
The 2021 version of the Isabelle theorem prover was released earlier this year
and supports the reconstruction of
Alethe proofs generated by veriT.  Users of Isabelle/HOL can invoke the \texttt{smt}
tactic. This tactic encodes the current proof goal as an SMT-LIB problem and
calls an SMT solver. Previously only the SMT solver Z3 was supported. Now
veriT is supported too. If the solver produces a proof, the proof is
reconstructed within the Isabelle kernel. In practice, users will seldom
choose the \texttt{smt} tactic themselves. Instead, they call the
Sledgehammer tool that calls external tools to find relevant facts. Sometimes,
the external tool finds a proof, but the proof cannot be imported into Isabelle,
requiring the user to write a proof manually.
The addition of the veriT-powered \texttt{smt} tactic halves~\cite{schurr-2021} the rate
of this kind of failures. The improvement is especially pronounced for proofs found by CVC4.
A key reason for this improvement is the support for the
conflicting-instance instantiation technique within veriT.  Z3, the
singular SMT solver supported previously, does not implement this
technique.  Nevertheless, it is Alethe that allowed us to connect veriT
to Isabelle, and we hope that the support for Alethe in other solvers
will ease this connection between powerful SMT solvers and other tools in the future.

The process of implementing proof reconstruction in Isabelle also
helped us to improve the proof format. We found both,
possible improvements in the format (like providing the Farkas' coefficient for
lemmas of linear arithmetic) and in the implementation (by identifying
concrete errors).
One major shortcoming of the proofs were rules that combined several simplification
steps into one.
We replaced these steps by
multiple simple and well-defined rules. In particular every simplification rule
addresses a specific theory instead of combining them.
An interesting
observation of the reconstruction in Isabelle is that some steps can be skipped
to improve performance.  For example, the proofs for the renaming of variables
are irrelevant for Isabelle since this uses De~Bruijn indices.
This shows that reconstruction specific
optimizations can counterbalance the proof length which is increased by fine-grained
rules.
We will take this prospect into account as we further refine the format.

\section{A Glance Into the Future}
The development of the Alethe proof format so far was not a monolithic
process.  Both practical considerations and research progress --- such as
supporting fine-grained preprocessing rules --- influenced the development process.
Due to this, the format is not fully homogeneous, but this approach allowed
us to quickly adapt the format when necessary. We will continue this
pragmatic approach.

\paragraph{Speculative Specification.}

We are writing a speculative specification.\footnote{The current
version is available at
\url{http://www.verit-solver.org/documentation/alethe-spec.pdf}.}
During the development of the Isabelle reconstruction it became necessary
to document the proof rules in a coherent and complete manner.  When we
started to develop the reconstruction there was only an automatically
generated list of rules with a short comment for each rule. While
this is enough for simple tautological rules, it does not provide a clear
definition of the more complex rules such as the linear arithmetic rules.
To rectify this, we studied veriT's source code and wrote an independent
document with a list of all rules and a clear mathematical definition
of each rule.
We chose a
level of precision for these descriptions that serves the implementer:
precise enough to clarify the edge case, but without the details that would make it a fully formal specification.
We are now extending this document to a full specification of the format.
This specification is speculative in the sense that it will not be cast
in stone. It will describe the format as it is in use
at any point in time and will develop in parallel with practical support for the
format within SMT solvers, proof checkers, and other tools.

\paragraph{Flexible Rules.}

The next solver that will gain support for the Alethe format is the upcoming
cvc5 solver. Implementing a proof format into another solver reveals where the
proof format is too tied to the current implementation of veriT. On the one
hand, new proof rules must be added to the format --- e.g., veriT does not
support the theory of bitvectors, while cvc5 does.
When CVC4 was integrated into Coq via a translation of its LFSC proofs into
Alethe proofs~\cite{SMTCoq}, an ad-hoc extension with bitvector rules was
made. A revised version of this extension will now be incorporated into the
upcoming specification of the format so that cvc5 bitvector proofs can be
represented in Alethe.
Further extensions to other theories supported by cvc5, like the theory of
strings, will eventually be made as well.

Besides new theories, cvc5 can also be stricter than veriT in the usage
of some rules.
This strictness can simplify the reconstruction, since less search is
required.
A good example of this is the \texttt{trans} rule that expresses
transitivity.
This rule has a list of equalities as premises and the conclusion is
an equality derived by transitivity.
In principle, this rule can have three levels of ``strictness'':
\begin{enumerate}
\item The premises are ordered and the equalities are correctly oriented (like in cvc5), e.g.,
  \(a=b\), \(b=c\), and \(c=d\) implies \(a=d\).
\item The premises are ordered but the equalities might not be correctly oriented (like in veriT), e.g.,
  \(b=a\), \(c=b\), and \(d=c\) implies \(d=a\).
\item Neither are the assumptions ordered, nor are the equalities oriented, e.g., \(c=b\),
  \(b=a\), and \(d=c\) implies \(d=a\).
\end{enumerate}

\noindent
The most strict variant is the easiest to reconstruct: a straightforward linear
traversal of the premises suffices for checking.
From the point of view of producing it from the solver, however, this version is
the hardest to implement. This is due to implementations of the
congruence closure decision procedure~\cite{Nelson1980,Downey1980} in SMT
solvers being generally agnostic to the order of equalities, which can lead to
implicit reorientations that can be difficult to track.
Anecdotally, for cvc5 to achieve this level of detail several months of work
were necessary, within the overall effort of redesigning from scratch CVC4's
proof infrastructure.
Since we cannot assume every solver developer will, or even should, undertake
such an effort, all the different levels of granularity must be allowed by the
format, each requiring different complexity levels of checking.

To keep the proof format flexible and proofs easy to produce, we will
provide different versions of proof rules, with varying levels of granularity as
in the transitivity example case above, by \emph{annotating} them.
This leverages the rule \emph{arguments},
which are already used by some
rules.
For example, the Farkas' coefficient of the linear arithmetic rule are provided
as arguments.
This puts pressure on proof checkers and reconstruction in proof assistants to
support all the variants or at least the most general one (at the cost of
efficiency).  Hence, our design principle here is that the annotation is
optional: the absence of an annotation denotes the least strict version of the
rule.

\paragraph{Powerful Tooling.}

We believe that powerful software tools may greatly increase the utility of a
proof format.
Towards this end we have started implementing an independent proof checker for
Alethe.
In contrast to a proof-assistant-based reconstruction, this checker will not be
structured around a small, trusted kernel, and correct-by-construction
extensions.  Instead, the user would need to trust the implementation does not
lead to wrong checking results.
Instead, its focus is on performance, support for multiple features and greater
flexibility for integrating extensions and refinements to the format.
The Isabelle checker is currently not suited
to this task --- one major issue is that it does not support SMT-LIB input
files.\footnote{A version capable of doing so was developed for Z3 but it was unfortunately lost.}

This independent checker will also serve as a proof ``elaborator''.
Rather than checking, it will also allow converting a coarse-grained proof,
containing implicit steps, to a fine-grained one, with more detailed steps. The
resulting proof can then be more efficiently checked by the tool itself or
via proof-assistant reconstructions.
An example of such elaboration is the transitivity rule.
If the rule is not in its most detailed version, with premises in the correct
order and none implicitly reordered, it can be elaborated by greedily reordering
the premises and adding proof steps using the symmetry of equality.
Note however that in the limit detailing coarse-grained steps can be as hard as
solving an SMT problem. Should such cases arise, the checker will rely on internal
proof-producing procedures capable of producing a detailed proof for the given
step. At first the veriT and cvc5 solvers, which can produce fine-grained proofs
for several aspects of SMT solving, could be used in such cases.

A nice side effect of the use of an external checker is that it could prune useless
steps. Currently SMT solvers keep a full proof trace in memory and print a pruned
proof after solving finishes. This is in contrast to SAT solvers that dump proofs
on-the-fly. For SAT proofs, the pruned proof can be obtained from a full trace by using a tool like
\textsc{Drat-Trim}. There is some ongoing work by Nikolaj Bjørner on Z3 to also generate proofs
on-the-fly, but it is not clear how to support preprocessing and quantifiers.\footnote{\url{https://github.com/Z3Prover/z3/discussions/4881}}

\section{Conclusion}
\label{sec:concl}

We have presented on overview of the current state of the Alethe proof format
and some ideas on how we intend to improve and extend the format, as well as
supporting tools.
In designing a new proof format supported across two solvers we hope to provide
a first step towards a format adopted by more solvers.
This format allows several levels of detail, and is thus flexible enough to reasonably
easily produce proofs in various contexts.  We intend to define a precise
semantics at each level though.
This distinguishes our format from other approaches, such as the TSTP
format~\cite{Sutcliffe2004}, that are probably easier to adopt but only
specify the syntax, leading to very different proofs generated by the various
provers supporting it.

One limit of our approach for proofs is that we cannot express global
transformations like symmetry breaking. SAT solvers are able to add clauses
(DRAT clauses) such that the overall problems is equisatisfiable. It is unclear
however how to add such clauses in the SMT context.

Overall, we hope to get feedback from users and developers to see what special
needs they have and exchange ideas on the proof format.

\paragraph{Acknowledgment}
We thank Bruno Andreotti for ongoing work on the proof checker and
Hanna Lachnitt for ongoing work on the cvc5 support.
We are grateful for the helpful comments provided to us by the anonymous
reviewers.
The first author has received funding from the European Research Council (ERC)
under the European Union’s Horizon 2020 research and innovation program (grant
agreement No. 713999, Matryoshka).
The second author is supported by the LIT AI Lab funded by the State of Upper Austria.

\bibliographystyle{eptcs}
\bibliography{generic}

\end{document}